\documentclass[preprint2]{aastex}
\usepackage{natbib}
\usepackage{epsf,epsfig}

\def\eg{{\it e.g., }}

\def\C{{\cal C}}
\newcommand {\simlt}{\lower.5ex\hbox{$\; \buildrel < \over \sim \;$}}
\newcommand {\simgt}{\lower.5ex\hbox{$\; \buildrel > \over \sim \;$}}

\begin{document}

\title{Multiple Peaks in the Angular Power Spectrum of the
Cosmic Microwave Background: Significance and
Consequences for Cosmology}

\author{
P.~de Bernardis\altaffilmark{1},
P.A.R.~Ade\altaffilmark{2},
J.J.~Bock\altaffilmark{3},
J.R.~Bond\altaffilmark{4},
J.~Borrill\altaffilmark{5},
A.~Boscaleri\altaffilmark{6},
K.~Coble\altaffilmark{7},
C.R.~Contaldi\altaffilmark{4},
B.P.~Crill\altaffilmark{8},
G.~De~Troia\altaffilmark{1},
P.~Farese\altaffilmark{7},
K.~Ganga\altaffilmark{9},
M.~Giacometti\altaffilmark{1},
E.~Hivon\altaffilmark{9},
V.V.~Hristov\altaffilmark{8},
A.~Iacoangeli\altaffilmark{1},
A.H.~Jaffe\altaffilmark{10},
W.C.~Jones\altaffilmark{8},
A.E.~Lange\altaffilmark{8},
L.~Martinis\altaffilmark{11},
S.~Masi\altaffilmark{1},
P.~Mason\altaffilmark{8},
P.D.~Mauskopf\altaffilmark{12},
A.~Melchiorri\altaffilmark{13},
T.~Montroy\altaffilmark{7},
C.B.~Netterfield\altaffilmark{14},
E.~Pascale\altaffilmark{6},
F.~Piacentini\altaffilmark{1},
D.~Pogosyan\altaffilmark{4},
G.~Polenta\altaffilmark{1},
F.~Pongetti\altaffilmark{15},
S.~Prunet\altaffilmark{4},
G.~Romeo\altaffilmark{15},
J.E.~Ruhl\altaffilmark{7},
F.~Scaramuzzi\altaffilmark{11}
}

\affil{
$^{1}$ Dipartimento di Fisica, Universita' La Sapienza,
        Roma, Italy \\
$^{2}$ Queen Mary and Westfield College, London, UK \\
$^{3}$ Jet Propulsion Laboratory, Pasadena, CA, USA \\
$^{4}$ Canadian Institute for Theoretical Astrophysics,
        University of Toronto, Canada \\
$^{5}$ National Energy Research Scientific Computing Center,
        LBNL, Berkeley, CA, USA \\
$^{6}$ IROE-CNR, Firenze, Italy \\
$^{7}$ Dept. of Physics, Univ. of California,
        Santa Barbara, CA, USA \\
$^{8}$ California Institute of Technology, Pasadena, CA, USA \\
$^{9}$ IPAC, California Institute of Technology, Pasadena, CA, USA \\
$^{10}$ Department of Astronomy, Space Sciences Lab and
        Center for Particle Astrophysics, \\
        University of CA, Berkeley, CA 94720 USA \\
$^{11}$ ENEA, Frascati, Italy  \\
$^{12}$ Dept. of Physics and Astronomy, Cardiff University, \\
        Cardiff CF24 3YB, Wales, UK \\
$^{13}$ Nuclear and Astrophysics Laboratory, University of Oxford,
        Keble Road, Oxford, OX 3RH, UK\\
$^{14}$ Depts. of Physics and Astronomy, University of Toronto, Canada \\
$^{15}$ Istituto Nazionale di Geofisica, Roma,~Italy \\
}

\begin{abstract}
Three peaks and two dips have been detected in the power spectrum of
the cosmic microwave background from the BOOMERANG experiment, at
$\ell \sim 210, 540, 840$ and $\ell \sim 420, 750$, respectively.  Using
model-independent analyses, we find that all five features are 
statistically significant and we measure their location 
and amplitude. These are
consistent with the adiabatic inflationary model. We also calculate
the mean and variance of the peak and dip locations and amplitudes in
a large 7-dimensional parameter space of such models, which gives good
agreement with the model-independent estimates, and 
forecast where the next few peaks and dips should be found if the
basic paradigm is correct.  We test the robustness of our results by
comparing Bayesian marginalization techniques on this space with
likelihood maximization techniques applied to a second 7-dimensional
cosmological parameter space, using an independent computational
pipeline, and find excellent agreement: $\Omega_{\rm tot} =
1.02^{+0.06}_{-0.05}$ {\it vs.} $1.04 \pm 0.05$, $\Omega_b h^2 =
0.022^{+0.004}_{-0.003}$ {\it vs.} $0.019^{+0.005}_{-0.004}$, and $n_s =
0.96^{+0.10}_{-0.09}$ {\it vs.} $0.90 \pm 0.08$. The deviation in
primordial spectral index $n_s$ is a consequence of the strong
correlation with the optical depth.

\end{abstract}

\section{Introduction}\label{sec:intro}

BOOMERANG has recently produced an improved power spectrum $\C_\ell$
of the cosmic microwave background (CMB) temperature anisotropy
ranging from $\ell \sim 100$ to $\ell \sim 1000$ \citep[][ hereafter
B01]{Nett2001}.  Three peaks are evident in the data, the first, at
$\ell \sim 210$ confirms the results of previous analysis of a small
subset of the BOOMERANG data set \citep[][ hereafter B00]{deBe2000} as
well as results from other experiments \citep[see
e.g.][]{miller1999,mauskopf2000,Hana2000}.  Analysis of the bulk of
the remaining data has improved the precision of the power spectrum
and extended the coverage to $\ell \sim 1000$, revealing for the first
time a second peak at $\ell \sim 540$, and a third at $\ell \sim 840$.

The results of two other experiments, released simultaneously 
with B01, are in good agreement with the  
BOOMERANG data. In addition to the first peak, 
the results from DASI \citep{Halv2001} show a 
peak coincident with that seen in our data near $\ell \sim 540$, and a rise 
in the spectrum toward high $\ell$ that is consistent with the leading 
edge of the peak seen in our data near $\ell \sim 840$.  The results from 
MAXIMA \citep{Lee2001} are of lower precision, but are consistent 
with both the DASI and the BOOMERANG results.

These detections are the first unambiguous confirmation of the
presence of acoustic oscillations in the primeval plasma before
recombination \citep{Peeb1970,SZ70}, as expected in the
standard inflationary scenario \citep{BE87}.

If the adiabatic cold
dark matter (CDM) model with power-law initial perturbations
describes our cosmogony, then the angular power spectrum of the CMB
temperature anisotropy is a powerful tool to constrain cosmological
parameters \citep[see e.g.][and references therein]{kamionkowski1999}.

In B01 a rigorous parameter extraction has been carried
out with the methods of \citet{Lange2000}, significantly improving the
constraints obtained from previous CMB analyses
\citep[see e.g.][]{dodelson2000,Melc2000,Lange2000,Balb2000,tegmark2000,jaffe2001,kmr,bond2000,bridle2000} on key cosmological parameters.

Furthermore, the new BOOMERANG spectrum gives  a value for the baryon 
fraction $\Omega_bh^2$ that is in excellent agreement with independent constraints 
from standard big bang nucleosynthesis (BBN), eliminating any hint of 
a conflict between the BBN and the CMB-derived values for the baryon 
density \citep[see e.g.][]{peeb2000}.

In this paper we support the conclusions of B01 by
presenting two complementary analyses of the measured power spectrum.
In Section~\ref{sec:power} we briefly compare these latest BOOMERANG results with
those of B00. In Section~\ref{sec:peaks}, using ``model-independent''
methods similar to that used in \citet{Knox2000}, we measure the
positions and amplitudes of the peaks in the spectrum. We also compute
the probability distribution of the theoretical power spectra used in
B01 given the measured $C_\ell$'s to 
estimate the averages and variances of the peaks and dips, as
in \citet{bond2000}.

In Section~\ref{sec:robust}, we extract the distribution of cosmological parameters
using a different grid of theoretical power spectra and a different
method for projecting onto one and two variable likelihoods than that
used in B01, and show good agreement between the two
methods.  The B01 $\C_\ell$-database used the cosmological
parameter set $\{\Omega_{tot}$, $\Omega_\Lambda$, $\Omega_b h^2$,
$\Omega_c h^2$, $n_s$, $\tau_C$, $\C_{10}\}$, where $n_s$ is the
spectral index of primordial density fluctuations, $\Omega_c$ and
$\Omega_\Lambda$ are the cold dark matter and vacuum energy densities
in units of the critical density, and $\C_{10}$ is an overall
normalization of the power spectrum.  The alternate grid described
here uses $\Omega_b$, $\Omega_m \equiv \Omega_c + \Omega_b$ and the
Hubble parameter $h$ instead of $\Omega_b h^2$, $\Omega_c h^2$ and
$\Omega_{tot}=\Omega_m+\Omega_\Lambda$, which are now derived
parameters. 

The determination of cosmological parameters is affected by the
presence of near-degeneracies among them \citep{Efst1999}. In
B01 and \citet{Lange2000}, this is improved through the
use of parameter combinations which minimize the effects of 
these degeneracies, except for the important one between $\Omega_{tot}$
and $\Omega_\Lambda$. However, as long as the
database is sufficiently extensive and finely gridded, one should
get nearly the same answer. In addition to using different parameter choices
to generate one and two
dimensional likelihood functions, we use likelihood maximization
rather than marginalization (integration) over the other variables.
Maximization could have a different (and often more conservative)
response to the presence of degeneracies in the model space.  We obtain 
excellent agreement between the two treatments, and also show that
when we apply our methods to the DASI data, we get the same results as
\citet{pryke2001}.  In Section~\ref{sec:disc} we report our conclusions.
\medskip

\section{The Power Spectrum: Comparison with Previous Results}\label{sec:power}

        The first results from the 1998/99 flight of BOOMERANG were 
reported one year ago by B00.  These included the 
first resolved images of the CMB over several percent of the sky, and 
a preliminary power spectrum based on analysis of a small fraction of the data. 
The power spectrum obtained  by B01 incorporates 
a 14-fold increase in effective integration time, 
and an $\sim 1.8$-fold increase in sky coverage.  Thus, the B01 
result obtains higher precision both at low $\ell$, where the 
precision is limited by sample variance, and at high $\ell$, where it is 
limited by detector noise.  Moreover, the B01 results 
make use of an improved pointing solution that has led to qualitative 
improvements in our knowledge of the physical beam and our 
understanding of the pointing jitter, which adds in quadrature with 
the physical beam to give the effective angular resolution of the 
measurement.

        We now understand that the pointing solution used in
B00 produced an effective beam size of $(12.7 \pm 2.0)$
arcmin, rather than the $(10 \pm 1)$arcmin assumed in that analysis,
due to an underestimate of the pointing jitter. 
The effect of
underestimating the effective beamwidth is to suppress the power
spectrum at small angular scales. However, this is not the reason
preventing the detection of a second peak from that dataset. 
In fact, when the effects of the pointing jitter are corrected for, 
the signal-to-noise ratio of the B00 spectrum 
at $\ell > 350$
is still insufficient to detect the second peak, and the data
are still compatible with flat bandpower. In
Fig.~1 we plot the spectrum of B00 corrected for the
jitter underestimate and for an overall gain adjustment ($+20\%$ in
$\C_\ell$, within the $\pm 20\%$ uncertainty assigned to
the absolute calibration in both B00 and B01).
The corrected  B00 spectrum is in excellent agreement 
with the B01 spectrum. 

\begin{figure*}[htb]
\centerline{\epsfig{file=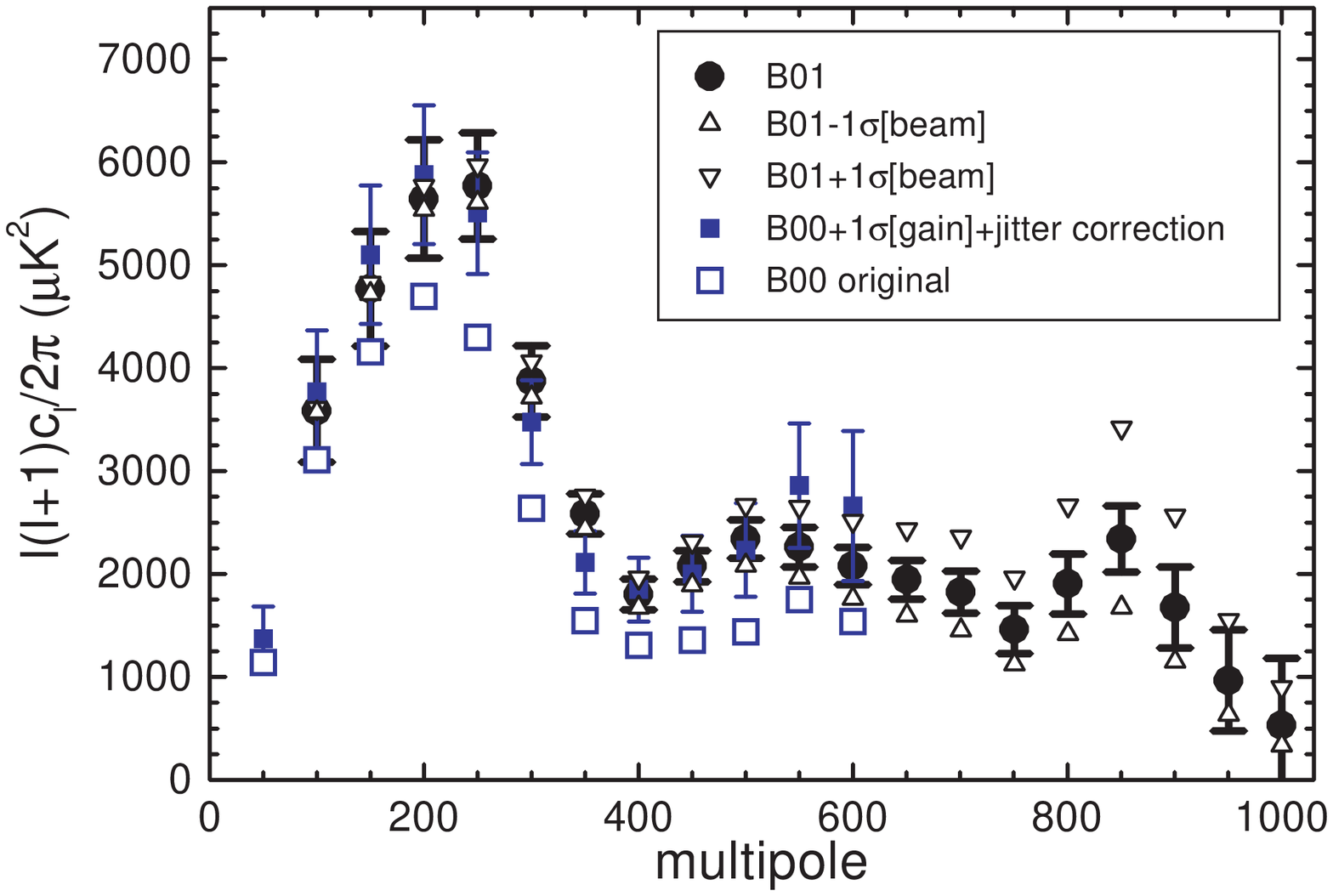,width=\textwidth}}
\caption{The angular power spectrum of the CMB measured by
  BOOMERANG.  The filled circles are the spectrum reported by B01; the
  up and down triangles indicate the systematic error in B01 resulting
  from the error in the beam.  Three peaks are visible at multipoles
  $\sim 210, \sim 540, \sim 840$.  The open squares are the spectrum
  reported by B00, based on an analysis of $\sim 7\%$ of the data
  analyzed in B01. The filled squares present the B00 spectrum after
  correction for pointing jitter and scaled by 10$\%$ ($1\sigma$) in
  overall calibration (see text).}
\end{figure*}

The data of B00 have been used to derive $\Omega_{tot}$
in the same paper, and to derive a full set of cosmological parameters
in \citet{Lange2000}. The cosmological result of B00 
- that the geometry of the universe is nearly flat - 
is not altered by the correction, since the correction does not 
move significantly the location of the first peak. 
That analysis was done assuming a "medium" prior on $h$ and $\Omega_b h^2$. 
When the full parameter analysis is done, assuming weaker priors 
as in \citet{Lange2000}, the effect of the correction is to drive $\Omega_{tot}$  
even closer to unity. With the new data set of B01, 
$\Omega_{tot} = 1.04^{+0.08}_{-0.07}$ using the $\ell < 600 $ part of the new data, 
$\Omega_{tot} = 1.02^{+0.06}_{-0.05}$ with all of it.

The largest effect of the correction is on the baryon density derived 
in \citet{Lange2000}. The uncorrected spectrum gives a baryon density
$\Omega_b h^2 = 0.036^{+0.006}_{-0.005}$ for the "weak" prior case.
The value derived from the B01 spectrum if only the data
at $\ell \le 600$ are used is $0.027^{+0.005}_{-0.005}$.  Adding the
data at higher $\ell$ removes degeneracies and further reduces the
value to $0.022^{+0.004}_{-0.003}$, as described below. 

\section{Significance and Location of the Peaks and Dips}\label{sec:peaks}

In this section we investigate the significance of the detection of
peaks and dips in our data, and estimate their locations using both
model-independent, frequentist methods and model-dependent, Bayesian methods.
\begin{deluxetable}{c|ccc|ccc}
\tablecaption{Peaks and dips: location and amplitude. \label{tab_peaks}}

\tablehead{
 & & \colhead{$\ell_{\rm p}$} & &  & \colhead{$\C_{\rm p}$ ($\mu
K^{2}$) } &
}

%
%
%
%
%
%
%
%
%
%
%
%
%
%
%


\startdata

Features & model independent & no priors & weak & model independent & no priors & weak \\

\tableline

Peak 1 & $213^{+10}_{-13}$        & $220\pm7$   & $221\pm5$   & $5450^{+1200}_{-1100}$ & $5440^{+1300}_{-1050}$ & $5140^{+1280}_{-1030}$ \\

Dip 1  & $416^{+22}_{-12}$  & $413\pm12$  & $412\pm7$   & $1850^{+440}_{-410}$  &$1640^{+390}_{-320}$    & $1590^{+390}_{-320}$   \\

Peak 2 & $541^{+20}_{-32}$ & $539\pm14$  & $539\pm8$   & $2220^{+560}_{-520}$ & $2480^{+620}_{-500}$   & $2420^{+630}_{-500}$   \\

Dip 2  & $750^{+20}_{-750}$ & $688\pm22$  & $683\pm23$  & $1550^{+550}_{-610}$ &$1530^{+470}_{-360}$    & $1590^{+500}_{-380}$   \\

Peak 3 & $845^{+12}_{-25}$ & $825\pm21$  & $822\pm21$  & $2090^{+790}_{-850}$ & $2170^{+720}_{-540}$   & $2270^{+760}_{-570}$   \\

Dip 3  &        -          & $1025\pm24$ & $1024\pm26$ &           -          &$880^{+290}_{-220}$     & $910^{+310}_{-230}$   \\

Peak 4 &         -         & $1139\pm24$ & $1138\pm24$ &           -          & $1100^{+380}_{-280}$   & $1130^{+400}_{-290}$   \\

Dip 4  &        -          & $1328\pm31$ & $1324\pm33$ &           -          &$570^{+220}_{-160}$     & $610^{+240}_{-170}$    \\

Peak 5 &         -         & $1442\pm30$ & $1439\pm32$ &           -          & $680^{+290}_{-200}$    & $730^{+310}_{-220}$    \\

Dip 5  &        -          & $1661\pm37$ & $1660\pm36$ &           -          &$300^{+130}_{-90}$      & $320^{+130}_{-90}$     \\

\tableline

\enddata

\centerline{\tablecomments{Location (columns 2-4) and amplitude (columns 5-7) of
peaks and dips in the power spectrum of the CMB measured by
BOOMERANG. In column 2 we list the values measured by means of a
parabolic fit to the bandpowers ($\Delta \ell = 50$). In columns 3 and
4 we list the values estimated by integrating the peak and dip
properties of the theoretical spectra used in \citet{Lange2000} and
B01 over the probability distribution for the database,
assuming either the no prior or weak prior restrictions of those
papers. In column 5,6,7 we do the same for the amplitude of the
peaks. Note that Peaks 4 and 5, and Dips 3, 4, 5 are all outside the
range directly measured, so they are forecasts of what is likely to
emerge if the database has components that continue to describe the
data well, as they do now. All the errors are at 1-$\sigma$ and 
include the effect of gain and beam calibration uncertainties. 
2-$\sigma$ confidence intervals can be significantly broader than
twice the 1-$\sigma$ intervals reported here, as is clear from Fig.~2.
}}
\end{deluxetable}\label{tab:table1}

\subsection{Model-Independent Analyses of Peaks and Dips}

\begin{figure}[htbp]
\epsfig{file=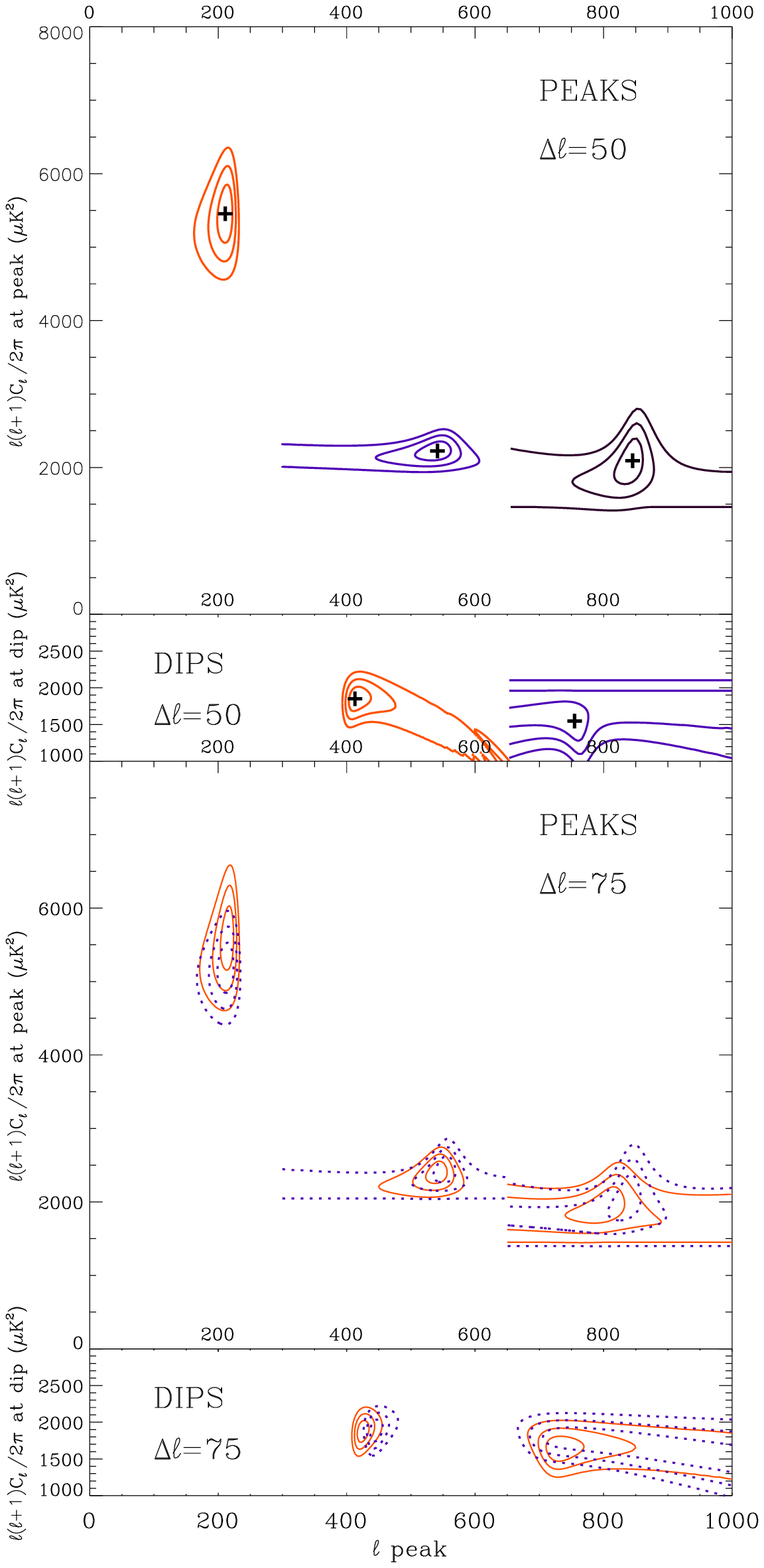,width=7cm}
\caption{\small $\Delta \chi^2$ contours for the position and
amplitude of the peaks and dips in the BOOMERANG CMB temperature power
spectrum. The crosses in the two upper panels correspond to the minima
for the $\chi^2$; the three contours are plotted at $\Delta \chi^2 =
2.3, 6.17, 11.8$ (corresponding to 68.3$\%$, 95.4$\%$, 99.7$\%$
confidence intervals for a gaussian likelihood in two parameters).
The two upper panels refer to a $\Delta \ell = 50$ binning of the
power spectrum. The multipole ranges used for the parabolic fits on
peak~1, dip~1, peak~2, dip~2 and peak~3 are (76 - 375), (276 - 475), (401 -
750), (676 - 875) and (726 -1025) respectively. The two lower panels
refer to binnings with $\Delta \ell = 75$. While the statistical
significance of the detections is improved, the location of the
features is less robust against the centering of the bins: continuous
and dashed lines refer to bin centers shifted by 35.}
\end{figure}

\subsubsection{Does a Flat Spectrum Fit the Data?}

As a first step, we answer the question, "How likely is it that the
measured bandpowers $\C_\ell$ are just fit by a first order polynomial
$\C_\ell^T = \C_A + \C_B \ell$".  Since the first peak is evident, we
limit this analysis to the data bins centered at $450 \simlt \ell
\simlt 1000$.  We compute $\chi^2 = (\C_b - \C_b^T) M^{-1}_{bb^\prime}
(\C_{b^\prime} - \C_{b^\prime}^T)$, where $M_{bb^\prime}$ is the
covariance matrix of the measured bandpowers $\C_b$ and $\C_b^T$ is an
appropriate band average of $\C_\ell$. For this exercise, we used a
Gaussian distribution in the $\C_b^T$, but have also checked a
lognormal in $\C_b^T$, with $M$ suitably transformed. We find similar
answers using these two limits of the offset-lognormal distribution
for the bandpowers recommended by \citet{BJK00}. We vary $\C_A$ and
$\C_B$ to find the minimum of the $\chi^2$, $\chi^2_{\rm min}$, and
compute the probability of having a $\chi^2 < \chi^2_{\rm min}$. A
first order polynomial is rejected in all cases.  $P(\chi^2 <
\chi^2_{\rm min})$ is 95.2$\%$, 96.5$\%$, 94.8$\%$ for the multipole
ranges 401-1000, 401-750, 726-1025 respectively (using a binning of
$\Delta \ell = 75$ for the bandpowers).  These conclusions are robust
with respect to variations in the location and width of the
$\ell$-ranges, as well as for variations of the beam FWHM allowed by
the measurement error.  We conclude that the features measured in the
spectrum are statistically significant at approximately $2\sigma$.

\subsubsection{Peak and Dip Location and Amplitude Likelihood Maps}

We now describe some methods we have used to locate the peaks and dips
of the power spectrum. Results shown in Table 1 used a
parabolic fit for $\C_\ell$. (As described below, we have also used
parabolas in $\ln \C_\ell$. The use of more complex functions, see
e.g. \citet{Knox2000}, is not required by the data.) The criterion for
robust peak/dip detection is that the mean curvature of the parabola
be above some threshold ({\it e.g.}, some multiple of the {\it rms}
deviation in the curvature). For the model-independent entries of
Table 1, for each amplitude $\C_{\rm p}$ and location
of the peak $\ell_{\rm p}$, we found the curvature which gave minimum
$\chi^2 (\C_{\rm p}, \ell_{\rm p})$. We repeat this procedure for
different data ranges. The contours corresponding to $\Delta\chi^2$ of
($2.3$,$6.17$ and $11.8$) are plotted in Fig.~2 for the ranges where
this procedure provides a detection of both $\C_{\rm p}$ and
$\ell_{\rm p}$. Three peaks and two dips are clearly detected by this
procedure, at the multipoles and with the amplitudes reported in the
table. The curvature is required to be negative for peak detection,
positive for dip detection. Note that zero curvature is equivalent to
the flat bandpower case. Approaching this zero curvature limit is why
the contours sometimes open up in $\ell_{\rm p}$ at $\sim 2\sigma$ and
become horizontal at $\sim 3\sigma$. This is a visual reiteration of
the point made above, that the flat $\C_\ell$ is rejected at about
2$\sigma$.  We have found that the location of the peaks is robust
against variations of the gain and beam calibration inside the
reported error intervals, $10\%$ for the gain calibration error,
producing a $20\%$ error in the $\C_{\rm p}$, and $1.4' (1\sigma)$
corresponding to $13\%$ in the beam.
The different best fit parabolas for the peaks
and dips of the BOOMERANG power spectrum are shown in Fig.~3.

\subsubsection{Fisher Matrix Approach with Sliding Bands}

We have tried a number of other model-independent peak/dip finding
algorithms on the data. For the parabolic model in $\ell$ for $\ln
\C_\ell$, we analyzed the quadratic form in blocks of 3, 4 or 5 bins,
each bin being of width $\Delta \ell =50$, determining the best fit
and the variance about it as defined by the inverse of the likelihood
curvature matrix (Fisher matrix). For this exercise, we adopted a
lognormal distribution of the bandpowers (but checked for robustness
by also assuming a Gaussian distribution) and marginalized over the beam
uncertainty. We slid across the $\ell$-space with our fixed bin group
and estimated the significance of the peak or dip detection by the
ratio of the best-fit curvature to the {\it rms} deviation in it.  For
BOOMERANG, using three bins of width $\Delta \ell =50$ gives stronger
detection, but results in larger error bars (as estimated from the
inverse Fisher matrix). Here, we quote our
5-bin results, and among these we use for each peak/dip  
the 5-bin template which gives 
the largest ratio of the mean curvature to  
standard deviation of the curvature. 



We find interleaved peaks and dips at $\ell =$ $215 \pm 11$, $431 \pm 10$,
$522 \pm 27$, $736 \pm 21$ and $837 \pm 15$. The error bars quoted
correspond to the variance in peak position $\ell_{\rm p}$ 
given the curvature of the parabola fixed at the best-fit value
( if one attempts to marginalize over the curvature, $\ell_{\rm p}$ loses
localization due to contribution from fits with vanishing curvature,
i.e. straight lines).
The amplitudes of the features are
$5760^{+344}_{-324}$, $1890^{+196}_{-178}$, $2290^{+330}_{-290}$,
$1640^{+500}_{-380}$ and $2210^{+900}_{-640}$ $\mu K^2$, correspondingly. When the 10\%
calibration uncertainty is included, the errors are similar to those
in Table 1, but of course the estimates move up and
down together in a coherent way, so here we have chosen to indicate
the purely statistical error bars. The significance of the detection
as estimated by the distinction of the best-fit curvature from zero is
$1.7\,\sigma$ for the second peak and dip, and $2.2\,\sigma$ for the third
peak.

\begin{figure*}[htb]
\centerline{\epsfig{file=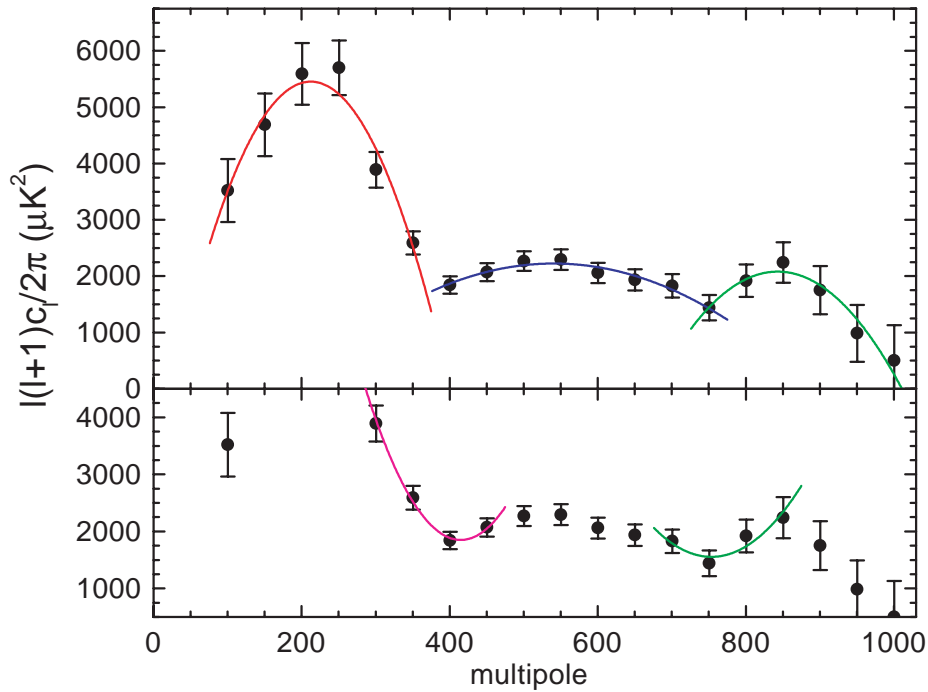,angle=-90,width=15cm}}
\caption{Best fit parabolas for the different 
peaks and dips of the BOOMERANG spectrum.}
\end{figure*}

\subsection{Model-Dependent Analysis of Peaks and Dips}

\subsubsection{Ensemble-Averaging Method}

We have also estimated peak/dip positions and amplitudes on a large
set of prescribed $\C_\ell$ shapes weighted by the probability that
each shape has when confronted with the data. For the cases shown in
Table 1, the prescribed shapes are the elements of the
$\C_\ell$ database used in \citet{Lange2000} and
B01. Explicitly, we compute for the BOOMERANG+DMR data the
ensemble averages $<\ln \ell_{\rm p} >$ and $<\ln C_{\rm p}> $ and
their variances, $<(\Delta \ln \ell_{\rm p})^2>$, and $<(\Delta \ln
C_{\rm p})^2>$, with respect to the product of the likelihood function
for the parameters and their prior probability. Two cases are shown,
for the no prior and weak prior cases of \citet{Lange2000}. Note the
good agreement using this method with the model-independent
results. (The results are nearly the same if we use $< \ell_{\rm p} >$
instead of $<\ln \ell_{\rm p} >$.)  When this method was applied to
all of the CMB data published before 2001, including that of BOOMERANG
and Maxima, a first peak location of $212\pm 7$ was obtained
\citep{bond2000}.

\subsubsection{Forecasting Peaks and Dips}

 One virtue of this procedure is that, because we know the spectral
shape for all $\ell$, we can forecast where peaks and dips will lie
beyond the $\ell$-region we observe. Thus, Table 1 gives predictions
for the locations and amplitudes of the subsequent (4,5) peaks and
(3,4,5) dips.  

To test whether this works using just the data at hand, we restrict
ourselves to using only the $\ell < 600$ data and see how well the
peaks beyond are found. As expected, the features with $\ell < 600$
are quite compatible with those in Table 1 derived
using all of the data: for the weak prior case, the first peak, the
first dip and the second peak for $\ell < 600$ are found to be at
$\ell =$ $219\pm 5$, $403 \pm 10$ and $530 \pm 23$, with amplitudes
$6000^{+1450}_{-1160}$, $1750^{+440}_{-350}$ and $2700^{+750}_{-590}$
$\mu K^2$. The second dip and third peak are out of range, predicted
to be found at $\ell =$ $640\pm 49$ and $789 \pm 40$, with amplitudes
$2080^{+790}_{-570}$ and $3160^{+1410}_{-980}$ $\mu K^2$, in good
agreement with the $\ell =$ $683\pm 23$ and $822 \pm 21$, with
amplitudes $1590^{+500}_{-380}$ and $2270^{+760}_{-570}$ $\mu K^2$
actually found. If we further restrict the prior probabilities, the
forecasted positions can move a bit: \eg we find the $\ell <600 $ cut
with the weak prior gives $\Omega_{tot}=1.04 \pm 0.07$, suggesting a
constraint to the theoretically-motivated flat universe is reasonable.
If we adopt as well the ``large scale structure'' prior of
\citet{Lange2000}, we get $\ell =$ $653\pm 45$ and $798 \pm 41$ at
amplitudes $1950^{+580}_{-450}$ and $2900^{+990}_{-740}$ $\mu K^2$;
and the fourth peak for $\ell<600$ is forecasted to be at $\ell = 1107
\pm 46$, similar to the $\ell = 1138 \pm 24$ forecast of the table.
Our conclusion is that even if we restrict ourselves to $\ell < 600$,
the forecasts are very good. We expect that using all of the data to
$\ell =1000$, with multiple peaks and dips, the forecasts of the Table
should be even more accurate.

\subsection{Peak and Dip Finding in the DASI data}

We have applied the sliding band procedure of Section 3.1.3 to the
DASI data using the 3-band and 4-band sliders. For the 3-band, we find
the first two peaks and interleaving dip at $\ell =$ $202 \pm 15$,
$407 \pm 14$ and $548 \pm 10$, with amplitudes $5200^{+530}_{-480}$,
$1590^{+163}_{-148}$ and $2760^{+270}_{-250}$ $\mu K^2$. Again, we
have not included the coherent 8\% calibration uncertainty (16\% in
$\C_\ell$) in these $\C_{\rm p}$ numbers. It may seems quite
incongruous that the error bars on the peak and dip positions can be
so small relative to the bin size, but we emphasize that these give
the error contours in the immediate neighborhood of the maximum, as
described by the Fisher matrix. Just as in Fig. 2, the contours
open up at levels lower than 1$\sigma$, resulting in imprecise
localization at the 2$\sigma$ level. DASI's detection of the
second dip is 1.6$\sigma$ in curvature by the Fisher error accounting,
but it is found,  at $\ell =$ $656 \pm 19$ and $\C_{\rm d} =
1650^{+260}_{-220} \mu K^2$.

We have also applied the prescribed shape method of Section 3.2.1 to the DASI data
\citep{Halv2001}.  We find that the location and forecasts are quite
compatible with those given in Table 1 for BOOMERANG. 
For example, for the weak
prior, the first three peaks and interleaved two dips are at $\ell =$ $216\pm
6$, $401 \pm 10$, $523 \pm12$, $655\pm 30$ and $794 \pm 30$, with
amplitudes $5240^{+1310}_{-1050}$, $1570^{+370}_{-300}$,
$2410^{+590}_{-470}$, $1690^{+410}_{-330}$ and
$2480^{+660}_{-520}$ $\mu K^2$. These $\C_p$ now include the $8\%$ calibration
uncertainty. The forecasted third dip is at $\ell = 988 \pm 34$ with
amplitude $980^{+260}_{-200} \mu K^2$, compatible with the BOOMERANG
result. 

\section{Robustness of Cosmological Parameters}\label{sec:robust}

The multiple peaks and dips are a strong prediction of the simplest
class of adiabatic inflationary models, and more generally of models
with passive, coherent perturbations \citep[e.g.][]{Albrecht:1996bg}. 
Although the main effect giving rise to them
is regular sound compression and rarefaction of the photon-baryon
plasma at photon decoupling, there are a number of influences that
make the regularity only roughly true. Nonetheless, the ``catalogue''
of peak and dip positions used to construct the table could be
searched to find best-fitting sequences and the associated
cosmological parameters giving rise to them. Indeed we know that peaks
and dips and only a few points in between are enough to characterize
the morphology of the $\C_\ell$ spectra
\citep[e.g.][]{sigscott00}.  However, it is clear that it is better to work
with the full shapes to test the theories. In this section, we show
that the extracted cosmological parameters using the full shapes are
robustly determined, by comparing results for the $\C_\ell$ database
and Bayesian marginalization techniques used in \citet{Lange2000} and
B01 with those obtained using the variables and likelihood
maximization techniques described below and in \citet{deBe1997},
\citet{dodelson2000}, \citet{Melc2000} and \citet{Balb2000}. We also test
our methods on the DASI data set.

\subsection{Extraction of Cosmological Parameters}

The $\{\Omega_{m},\Omega_{b},\Omega_{\Lambda},h,n_s, \tau_C ,
C_{10}\}$ space has parameters sampled as follows: $\Omega_{m}= 0.11,
..., 1.085$, in steps of 0.025; $\Omega_{b} = 0.015, ...,0.20$, in
steps of 0.015; $\Omega_{\Lambda}=0.0, ..., 0.975$, in steps of 0.025;
$h=0.25, ..., 0.95$, in steps of 0.05; spectral index of the
primordial density perturbations $n_s=0.50, ..., 1.50$, in steps of
0.02, $\tau_C = 0., .., 0.5$, in steps of 0.1.  The overall amplitude
$\C_{10}$, expressed in units of $\C_{10}^{COBE}$, is allowed to vary
continuously.  The theoretical models are computed using the {\sc
  cmbfast} program \citep{sz}, as were those used in B01.
Here, we have ignored the role gravity waves may play.  
We used the new BOOMERANG anisotropy power spectrum expressed as $18$
bandpowers from $\ell=70$ to $\ell=1050$ (see B01) and
we computed the likelihood for the cosmological models as $\exp(-\chi^2 /2)$,
where $\chi^2$ is the quadratic form defined in section {\it 3.1.1}. 
A $10 \%$ Gaussian-distributed calibration error in the
gain and a 1.4' (13\%) beam uncertainty were included in the analysis.
The COBE-DMR bandpowers used were those of \citet{BJK98}, obtained
from the RADPACK distribution \citep{Radpack}. For the other $\C_\ell$
database in which $\Omega_k = 1- \Omega_{tot}$ replaces $\Omega_m$ and
$\Omega_{c}h^2$ and $\Omega_{b}h^2$ replace $h$ and $\Omega_b$, the
parameter grid, the treatment of beam and calibration uncertainty, the
use of the offset-lognormal approximation, and the marginalization
method, are as described in \citet{Lange2000}.

As discussed in \citet{Lange2000}, apart from the inevitable database
discreteness, assuming a uniform distribution in the variables of the
database is tantamount to adopting different relative prior
probabilities on the variables. Of course, this is not an issue for
the maximization method, and even for marginalization is a very weak
prior relative to the strong observed detections. We can use further
``top hat'' priors to mimic the $\{\Omega_{m},\Omega_{b}, h\}$
database by restricting the wide coverage we have in the
$\{\Omega_{k}, \Omega_{b}h^2, \Omega_{c}h^2 \}$ variables. We
illustrate this in Fig.~4, where this mimicking prior is coupled to a
weak cosmological prior, requiring the age of the Universe to be above
10 Gyr and the Hubble parameter to be in the range 0.45 to 0.85. This
is very similar to the weak prior of \citet{Lange2000},
\citet{jaffe2001} and B01, except the upper limit was 0.95:
this results in only tiny differences in the extracted parameters.

\subsection{Cosmological Parameter Results}

Our basic results, on method testing and cosmological implications,
are shown in Fig.~4 and the associated contour plots Figs.~5 and~6.
Fig.~4 clearly shows that it does not make that much difference in
constructing the likelihood function for a target cosmological
variable if we marginalize (integrate) over the other cosmological
parameters or find the maximum likelihood value. Nor does it matter
which database is used. Similar success in agreement is found with
other prior choices, but for this paper we will restrict ourselves to
the $0.45 < h < 0.85$ weak prior. When we integrate the distributions of
Fig.~4 to get the 50\% value and the 1-sigma errors derived from the
16\% and 84\% values, we obtain $\Omega_{tot} = 1.02^{+0.06}_{-0.05}$,
$1.02^{+0.05}_{-0.06}$, $1.04 \pm 0.05$, for marginalization and
maximization in the \citet{Lange2000} $\C_\ell$-database and the
maximization in the alternate database, respectively; we also obtain
$n_s = 0.96^{+0.10}_{-0.09}$, $0.90^{+0.09}_{-0.09}$, $0.90 \pm 0.06$,
and $\Omega_b h^2 =0.022^{+0.004}_{-0.003}$,
$0.020^{+0.004}_{-0.004}$, $0.019^{+0.005}_{-0.004}$.

We have also applied the B01 $\C_\ell$ database with
marginalization to the DASI data, and find for our $0.45 < h < 0.95$,
age $> 10$ Gyr, prior $\Omega_{tot} = 1.05^{+0.06}_{-0.06}$, $n_s =
1.02^{+0.10}_{-0.09}$ and $\Omega_b h^2 =0.024^{+0.005}_{-0.004}$, to
be compared with the \citet{pryke2001} values of
$1.05^{+0.06}_{-0.06}$, $1.01^{+0.09}_{-0.07}$ and
$0.022^{+0.004}_{-0.004}$. As mentioned above, the overlap of 20\% of
the DASI fields one the sky precludes a rigorous joint analysis of the
datasets. If you proceed anyway as an exercise and ignore those 
correlations,  the results are very close to the BOOMERANG+DMR 
marginalization values given above,
with very slightly reduced errors.

\begin{figure*}[htb]
\centerline{\epsfig{file=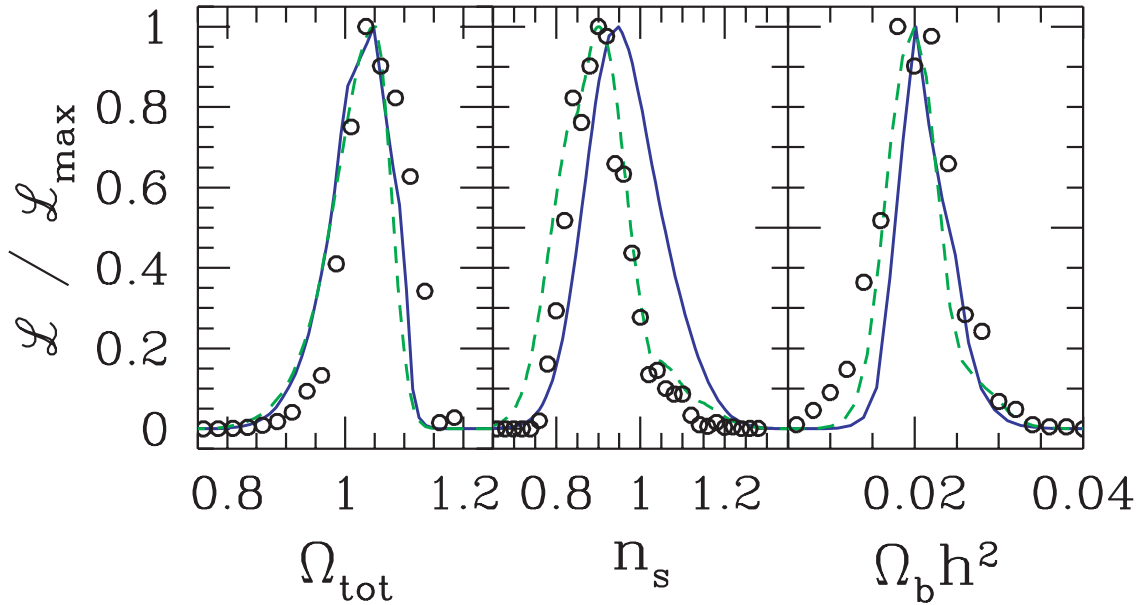,width=15cm}}
\caption{Likelihood curves derived from the BOOMERANG
and COBE/DMR datasets as a function of $\Omega_{tot}$, $n_s$ and
$\Omega_b h^2$ show the relative insensitivity to whether
marginalization over the other variables is done (solid) or the
maximum likelihood point in the other variables is chosen
(dashed). These curves were constructed using a prior on the
$\C_\ell$-database used in B01 designed to mimic the
ranges in the alternate $\C_\ell$-database that uses $\{ \Omega_b, 
\Omega_m,h\}$ (open circles). A further weak
prior, with $0.45<h<0.85$ and age $> 10$ Gyr, was used. The high
degree of correlation of $n_s$ with $\tau_C$ accounts for the wider
distribution of the marginalized $n_s$ likelihood compared to that for
the maximization procedure.}
\end{figure*}

\begin{figure}[htb]
\centerline{\epsfig{file=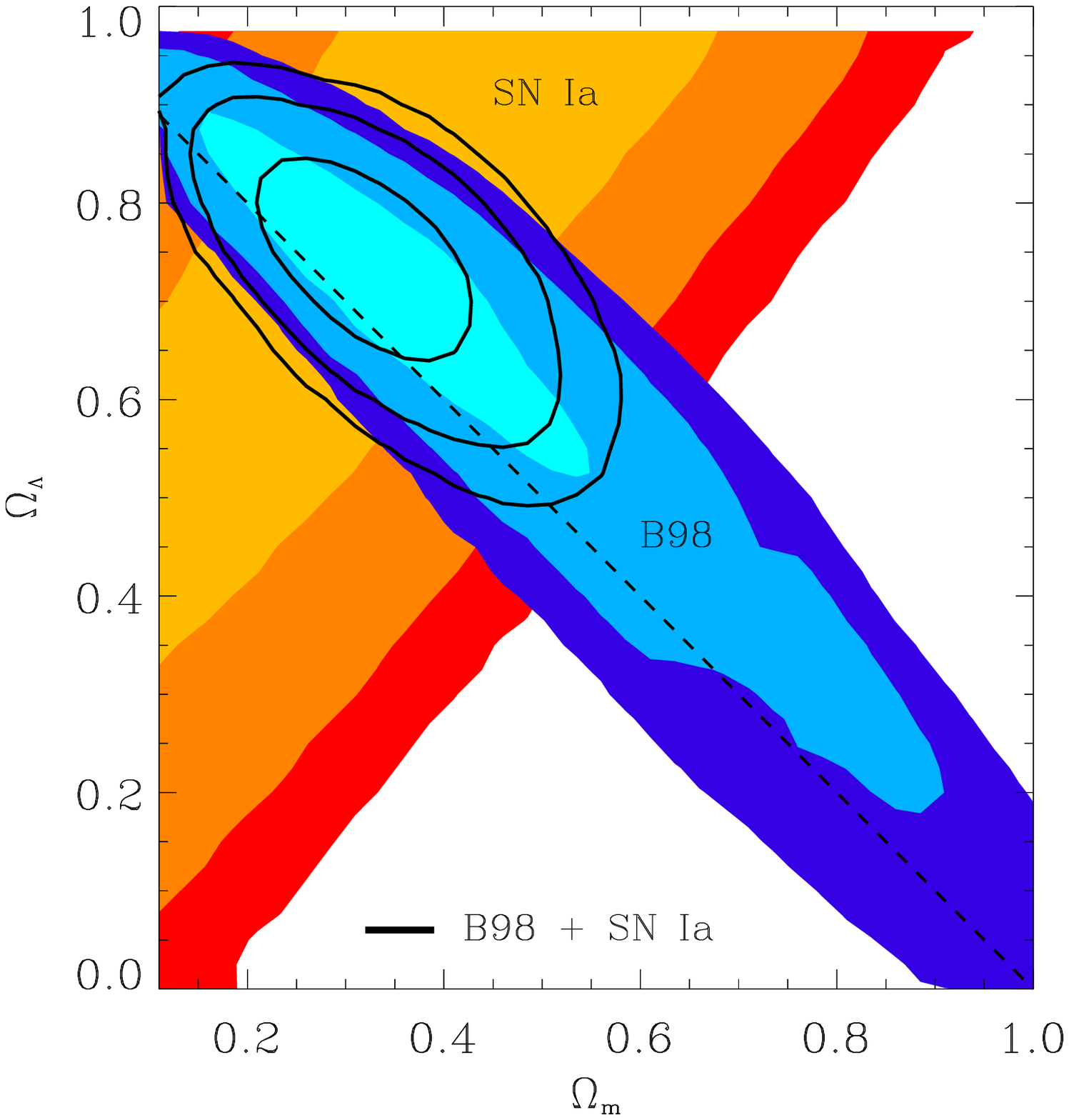,width=8.5cm}}
\caption{Constraints in the
$\Omega_m$ vs. $\Omega_\Lambda$ plane from the combined BOOMERANG and
COBE/DMR datasets, assuming the weak prior  $0.45 < h <
0.85$, age $> 10$ Gyr. For this plot, the likelihood at each point is
calculated by maximizing over
the remaining $5$ parameters. The shaded regions correspond to
the $68.3\%$, $95.4\%$ and $99.7\%$ confidence levels. The CMB
contours (labeled B98) are overlaid on the contours derived from
observations of high redshift supernovae (labeled SN 1a). The line
contours are for the combined likelihood.}
\end{figure}

We now discuss the implications of these determinations in more
detail.  If we knew all other variables, the position of the first
peak would allow us to determine the mean curvature of the universe,
and all subsequent dips and peaks would have to follow in a specific
set pattern. However, degeneracies among cosmological parameters are
present. In particular, there is a geometrical degeneracy between
$\Omega_\Lambda$ and $\Omega_{tot}$ in the angular diameter distance,
which allows the peak/dip pattern to be reproduced by different
combinations of the two. The peak/dip heights also have
near-degeneracies associated with them, but these can be strongly
broken as more peaks and dips are added. Extreme examples are closed
models dominated by baryonic dark matter, which can reproduce the
observed position of the first peak \citep{grif2001} but are unable
to account for the observed $\ell > 600$ power.

The classic $\Omega_m$ vs. $\Omega_\Lambda$ plot in Fig.~5 shows $68.3\%$,
$95.4\%$ and $99.7\%$ contours, defined to be where the likelihood
falls to $0.32$, $0.05$ and $0.01$ of its peak value, as it would be
for a 2D multivariate Gaussian.  The phase space of models in the
$\Omega_m$ vs. $\Omega_\Lambda$ plane results in the likelihood for the
total energy density of the universe $\Omega_{\rm tot}$ being skewed
towards closed models \citep[see e.g.][]{Lange2000,bond2000}.
This skewness is not important if the acceptable model space is well 
localized, which can be accomplished by imposing 
prior probabilities, in particular on $h$. The weak ``top hat $h=0.65 \pm
0.20$'' prior we have adopted decreases the skewness to closed models
(which is quite evident if models with very low $h$ values are
included, as described in B01, and helps to break the
geometrical degeneracy, hinting at the presence of a cosmological
constant at the level of $\sim 1 \sigma$. This becomes more pronounced
with a stronger prior on $h$ (B01).  Including the recent
supernovae data \citet{super1} with the weak prior used here, we find
$\Omega_\Lambda=0.71\pm 0.11$ and
$\Omega_m=0.31^{+0.13}_{-0.12}$ using maximization, to be contrasted
with the marginalization results using the B01 database of
$0.73^{+0.07}_{-0.10}$ and $0.32\pm 0.06$.

The measurement of the relative amplitude of the peaks in the CMB
spectrum provides important constraints on the physical density of
baryons $\Omega_b h^2$.  In the region from $\ell \sim 50$ and up to
the second peak, $\Omega_b h^2$ is nearly degenerate with variations
in the primordial spectral index of scalar fluctuations $n_s$:
increasing $\Omega_b h^2$ increases the ratio of the amplitudes of the
first and second peaks, but so does decreasing $n_s$.  Beyond the
second peak, however, the two effects separate: the third peak rises
by increasing $\Omega_b h^2$ and lowers by decreasing $n_s$. Thus,
though Fig.~6 shows that the two are still well-correlated, the
inclusion of BOOMERANG data up to $\ell \sim 1000$ has sharpened our
ability to independently estimate the two.

Of course, both $\Omega_b h^2$ and $n_s$ are extremely important for
our understanding of the early universe. The $\Omega_b h^2$ value can
be inferred from observations of primordial nucleides under the
assumption of standard BBN scenarios.  Recent observations of
primordial deuterium from quasar absorption line systems suggest a
value $\Omega_b h^2=0.020 \pm 0.002$ at the $95 \%$
C.L. \citep{burl2000}.

In inflation models, the spectral index of the primordial fluctuations
gives information about the shape of the primordial potential of the
inflaton field which drove inflation. While there is no fundamental
constraint on this parameter, the simplest and least baroque models of
inflation do give values that are just below unity.  We obtain 
$n_s=0.90\pm 0.09$ using the maximization procedure
and the $n_s=0.96\pm 0.09$ using the preferred marginalization
method.

\begin{figure}[htb]
\centerline{\epsfig{file=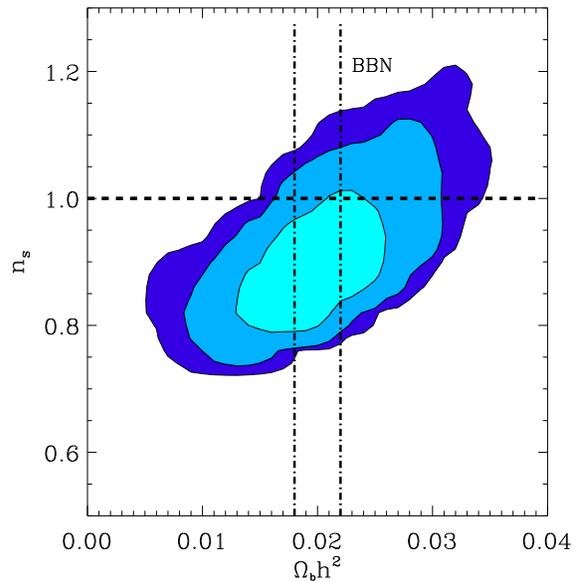,width=8.5cm}}
\caption{Constraints in the $\Omega_bh^2$ vs. $n_s$
plane from the combined BOOMERANG and COBE/DMR datasets. The
likelihood at each point is computed by maximizing over the remaining
five parameters. The shaded regions correspond to $68.3\%$, $95.4\%$
and $99.7\%$ confidence levels. The bound obtained from standard
BBN is overlaid. }
\end{figure}

Since $n_s$ and $\tau_C$, the optical depth to the surface of last
scattering, are also highly correlated, there is a noticeable
difference between marginalization and maximization, as seen in
Fig.~4.  Increasing $\tau_C$ suppresses the high-$\ell$ power spectrum
by a constant factor $\sim \exp{-2 \tau_C}$, but leaves the lowest
multipoles unaffected. This is similar to the effect of altering the
spectral index, which changes higher multipoles with a longer ``lever
arm'' with respect to the lower $\ell$. Because the likelihood
functions peak for nearly all models at $\tau_C=0$, the maximization
procedure effectively ignores $\tau_C>0$, whereas marginalization
averages over the allowed values of $\tau_C$, accounting for the
increase from 0.90 to 0.96. (If we restrict ourselves to $\tau_C=0$,
we get $0.89^{+.09}_{-.08}$ with maximization, $0.92^{+.08}_{-.07}$
with marginalization.)

The values of the spectral index, the curvature and the cosmological
constant affect the shape and the amplitude of the power spectrum of
the matter distribution.  The bandpower $\sigma_8^2$, giving the
variance in (linear) matter fluctuations averaged in $8 h^{-1} Mpc$
spheres, is often used to characterize the results of many large scale
structure observations. In \citet{Lange2000}, \citet{jaffe2001} and
B01, $ \sigma_8
\Omega_m^{0.56}$=$0.55^{+.02,+.11}_{-.02,-.08}$ was adopted, where the
combination of Gaussian and top hat error bars were used to generate a
wide distribution ``weak'' large scale structure prior to be imposed
upon the CMB data. This compares with $\sigma_8\Omega_m^{0.5} \sim
0.50\pm0.05$ estimated by \citet{pen1998} and a best fit value
$\sigma_8\Omega_m^{0.47} \sim 0.56$ from \citet{viana1999}.  We have
computed $\sigma_8$ for each model in our database using the matter
fluctuations power spectrum. From the likelihood analysis in the space
($\sigma_8, \Omega_m$), we find the 1-sigma range $0.5 \simlt \sigma_8
\simlt 0.8$ for $0.3 \simlt \Omega_m \simlt 0.7$. For the detailed
application of the weak large scale structure prior
of \citet{Lange2000}, which also included a constraint on the shape of
the power spectrum, see B01.

\section{Discussion and Conclusions}\label{sec:disc}

        We have analyzed the most recent results from BOOMERANG, and 
shown that there are a series of 5 features, 3 peaks with 2 
interleaved dips, each of which is detected using a cosmological 
model independent method at $\sim 95 \%$ or better confidence.  Because 
BOOMERANG is able to achieve resolution in $\ell$-space of 
$\Delta \ell \sim 50$, with only $\sim 10\%$ 
(anti)-correlation between bins, the data determine 
the positions as well as the amplitudes of each of the features with 
reasonable precision.

        The positions and amplitudes of the features are 
consistent with the results of the DASI experiment. 
A direct, cosmological model-independent 
comparison of the two experiments is made difficult by the lower 
$\ell$-space resolution of the DASI experiment.   However, the two results 
can be accurately compared via a model-dependent analysis that takes 
millions of $\C_\ell$ shapes with very different peak/dip amplitudes and
locations, weighted their peak and dip locations with the probability
of the (normalized) shapes, and finds well-localized $\ell_{\rm
p}$ and $\C_{\rm p}$ for each of the peaks and dips. The agreement between
BOOMERANG and DASI using this method is very good, and
completely consistent with the model-independent results. It will of
course be of great interest whether the forecasts allowed by this
model-dependent technique for the positions and amplitudes of the next
peaks and dips will be borne out by even higher resolution
observations.

The natural interpretation of this much-antici\-pated sequence of peaks
and dips in $\C_\ell$ is that we are seeing phase-coherent pressure
waves in the photon-baryon fluid at photon decoupling at redshift
$\sim 1100$, expected in adiabatic models of structure formation, and
which therefore the simplest inflation models give. This is certainly
the most economical interpretation, especially since the locations
within this paradigm correspond to widely-anticipated cosmological
parameter choices, namely $\Omega_{tot}\approx 1$, $\Omega_m \sim
1/3$, $\Omega_Bh^2 \approx 0.02$, with a nearly scale invariant
spectrum, $n_s \sim 1$. We have shown how well determined and robust
these parameter values are to changes in the $\C_\ell$ database, in
using likelihood maximization or Bayesian marginalization, and to using
either the BOOMERANG or DASI $\C_\ell$s in conjunction with the DMR
$\C_\ell$s. 

Of course this does not clinch the case: the derived cosmological
parameters could be quite different if we allowed ourselves much
further freedom beyond a single slope to characterize the primordial
spectrum. It could even be that the peaks and dips reflect early
universe structures rather than sound wave structures. It is also
possible that isocurvature modes with artful enough initial condition
choices could mimic the simple adiabatic case
\citep[e.g.][]{turok1996}. The detection of the associated polarization
peaks and dips would rule out these more exotic
possibilities. Although there are no glaring anomalies between theory
and data at this stage requiring a revisit of basic CMB assumptions,
the improved precision in this $\ell$-range that more sky coverage
will give, and the extension to higher $\ell$ that other CMB
experiments will give, could well reveal that our forecasts are
wrong. Even with the current data, there is certainly room for other
cosmological parameters not treated in our minimalist inflation
databases, and such explorations are underway.

\acknowledgments

The BOOMERANG project has been supported by the CIAR and NSERC in
Canada, by PNRA, Universit\'a ``La Sapienza'', and ASI in Italy, by
PPARC in the UK, and by NASA, NSF OPP and NERSC in the U.S.  We
received superb field and flight support from NSBF and the USAP
personnel in McMurdo.

\end{document}